\documentstyle[preprint,prc,aps]{revtex}
\begin{document}
\preprint{
\rightline{\vbox{\hbox{\rightline{MSUCL-978}} 
\hbox{\rightline{nucl-th/9604035}}}}
         }
\title{Excess electron pairs from heavy-ion collisions at CERN
and a more complete picture of thermal production}
\author{Kevin L. Haglin\cite{myemail}}
\address{
National Superconducting Cyclotron Laboratory, Michigan
State University\\
East Lansing, Michigan 48824--1321, USA}
\date{\today}
\maketitle 
\begin{abstract}
The low-mass dielectron signal from heavy-ion collisions at the CERN--SPS
reported by the {\em CERES} collaboration is in excess of estimated
hadronic decays suggestive of possible contribution from two-pion annihilation
or other hadronic reactions.  In the absence of dramatic medium 
modifications, annihilation alone is unable to account 
for the data.   We explore the role of pion plus resonance scattering 
[$\pi\rho\to a_{1}(1260) \to \pi e^+e^-$] which has favorable
kinematics to populate masses between $2m_{\pi}$ and $m_{\rho}$.
While it seems to account for some of the remaining excess beyond 
annihilation, it fails to allow quantitative interpretation of data.
\end{abstract}
\pacs{PACS numbers: 25.75.+r,13.40Ks}

\narrowtext

Dynamics of high-energy heavy-ion collisions can be probed, at least in
principle, by photons and dileptons since they are produced continuously
from the first moments of nuclear contact throughout the entire evolution
and interact only weakly through electromagnetic coupling to the hadronic
medium.  Their mean free paths are many times greater than typical
reaction-zone sizes allowing them to bring out valuable information 
on the earlier stages.  Since their utility for heavy-ion collisions was 
first proposed\cite{es80}, many theoretical 
calculations have been done\cite{hwa85,kk86,jcl91,kh92} predicting
orders of magnitude for production, exploring ideas about phase transition 
to quark matter and possible medium modifications to hadronic phenomena.   
Recently an intriguing report from CERN\cite{ga95} brought forth results for 
dielectron signals observed in heavy-ion induced reactions as compared with 
proton-induced results.   While the p+Be and p+Au data at 450 GeV were
not inconsistent with estimations from hadronic decays, the S+Au results
showed an enhancement by a factor of 5.0$\pm$ 2.7 over hadronic decay
contributions when integrated over mass from 0.2 to 1.5 GeV.
                                                               
The natural conclusion drawn from these theoretical estimates and experimental
data was that the excess is most probably due to $\pi\pi$ annihilation 
in a heated hadronic environment.  But since the excess is
pronounced for masses near 500 MeV and lower, this is not 
obviously annihilation.  If not, then what?  Perhaps we are seeing the
rho mass shifting downward by several hundred MeV due to the presence of the
medium\cite{li95}; or perhaps we are seeing effects of modified pion dispersion 
relations\cite{koch95}.  Both phenomena would be spectacular if only
we could unequivocally establish their influences which at this stage
are matters of discussion.  In fact, prediction for direction and 
magnitude of mass shifts at finite energy density are not yet settled as 
the literature bears opposing results\cite{brown,pisarski}.  Taking the 
point of view to use free
space masses and widths (and including a collision broadening for $\rho$), 
we show in this letter that the ``excess'' dielectron signal is due in 
part due to $\pi\pi$ annihilation, but could have important 
contributions from pion-resonance $\pi\rho\to\pi e^{+}e^{-}$ scattering.

Consider a central collision of sulfur on gold at 200 GeV/u which will
produce a ``fire cylinder'' of transverse radius $R_{T}$ equal to that of
the sulfur nucleus.  The question of initial energy densities and QCD phase
transition are quite interesting but will not be discussed in any detail
here.  Only to note that assuming isentropic expansion and relating observed
rapidity densities to initial times and temperatures\cite{hwa85}, $T_{i}$ 
is quite high---of the order 350 MeV.  Instead, if either the mixed phase 
(of quark and hadron coexistence) or pure plasma phase is reached, the 
additional degrees of freedom will effectively keep the temperature
lower.  This is the picture envisioned so that $T_{i}$ is of the order
180 MeV.  Simulations of relativistic transport\cite{li95} support
this picture since they are able to reproduce transverse mass spectra 
for both pions and protons with an initial temperature of $\sim$ 185 MeV
and a finite chemical potential $\mu_{\pi}=135$ MeV.  Being only
somewhat above the critical value of temperature obtained from lattice 
calculations\cite{christ} for entering the plasma phase, we neglect
contributions from quark processes.  The system expands and cools very
quickly until the mixed phase is reached when hadronic contributions begin
thermally producing dielectrons at $T_{c}$ for a few fm/$c$ until the matter
is found to be in a complete hadronic phase.   It cools and expands further
until collisions can no longer support the nearly isotropic momentum 
distributions and freezes out.

The observed spectrum of dielectrons will consist of several components,
each coming from different mechanisms and possibly different stages of the
evolution.  Below 150 MeV mass a nonthermal component of $\pi^{0}$ and 
$\eta$ Dalitz dominates and the observed yields are consistent with 
estimates\cite{ga95}.  Hadronic radiative decays will contribute 
to higher masses and the largest seems to be $\omega\to \pi^{0}e^{+}
e^{-}$\cite{ga95,cgpl94}, although there are recent suggestions that
partial $U(1)_{A}$ restoration could modify $\eta^{\prime}$ physics in
ways relevant for this experiment\cite{kapusta} and boost 
$\eta^{\prime}$ Dalitz to a more important level.
Direct decay of vector mesons $\rho^{0}$ and
$\omega$ will also be a strong source of pairs but fairly near their 
free space masses.  Production of a nonzero four volume of 
hot matter generates contribution from pion annihilation 
$\pi^{+}\pi^{-}\to e^{+}e^{-}$, whose strength depends on the average 
temperature.  In this four volume there are myriad hadronic reactions 
able to produce pairs.
A systematic treatment of decays and annihilation, i.e. lowest order 
treatment in the strong coupling has been done\cite{cgpl94}.  Going beyond 
lowest order is required for this discussion as $\pi\rho\to\pi\,e^{+}e^{-}$
comes from a two-loop contribution.   Elastic scattering
of $\pi$'s with $\rho$'s is dominated by resonant $a_{1}$ formation as
the average $\sqrt{s}$ for $T\sim 150$ MeV is near the centroid
of the rather broad $a_{1}$ mass distribution.
Since the $a_{1}$ has a radiative decay channel as well, it follows
that scattering can produce real or virtual photons.  For real
energetic photon production, the reaction $\pi\rho\to a_{1}\to\pi\gamma$ has
been shown to be very important\cite{lxesgb,song}.  As $E_{\gamma}\to 0$
phase space shrinks to a point and this mechanism
contributes nothing.  Zero invariant mass for virtual photons does not 
require zero energy as $q_{0}^{2}$ and $\vec{q}^{\,\,2}$ could separately
be very large.  Consequently the amplitude for $\pi\rho\to\pi\,e^{+}e^{-}$ will 
go like $1/M^{2}$ for small mass.  Near the rho, vector dominance takes 
over and the amplitude goes like $1/(M^{2}-m_{\rho}^{2}+im_{\rho}
\Gamma_{\rho})$.  A completely equivalent and alternative approach to 
estimating the role of the $a_{1}$ as considered here would be to compute 
$a_{1}\to \pi e^{+}e^{-}$ starting from a thermal distribution of charged
$a_{1}$'s properly modified by a Breit-Wigner distribution in 
mass\cite{weldon93} owing to its finite width.

More general techniques of field theory at finite temperature unambiguously
determine the total contribution\cite{weldon,jk}.  The rate for producing 
electron pairs is related to the imaginary part of the retarted photon 
self-energy which, to one-loop order, corresponds to decay and annihilation.  
The pion-resonance scattering process being of two-loop order would require
additional renormalization and the added nuisance of an additional Matsubara
sum, but the prescription is clear.  As the field theoretic methods are 
completely equivalent to kinetic theory at given order in the coupling, the 
latter is chosen for convenience.  It builds upon
evaluating tree-level Feynman diagrams and folding thermal momentum 
distributions with transition amplitudes to arrive at an average rate in 
medium for a given process.  For instance, the dominant $a_{1}$ resonant 
graph contributing to $\pi\rho\to\pi e^{+}e^{-}$ is shown in
Fig.~\ref{fig:feynman}.  Interference and other meson-exchange effects
will be reported upon separately\cite{forthcoming}.  The appearance 
of the intermediate rho suggests 
using a vector-dominance form factor whose width is taken to be a function of 
temperature to include collision broadening. It is computed just as in 
Ref.~\cite{khcoll95}
but including finite pion chemical potential and then parametrized for
100 $< T <$ 200 MeV by
\begin{eqnarray}
\Gamma_{\rho}(T) &=& \Gamma_{\rho} + \left[a+b\,T+c\,T^{2}\right]
\end{eqnarray}
where $\Gamma_{\rho}$ is the free space width and $a=0.50$ GeV,
$b=-7.16$ and $c=30.16$ GeV$^{-1}$.

An effective lagrangian for axial-vector--vector--pseudoscalar interaction
is chosen for simplicity to be the following\cite{lxesgb}
\begin{eqnarray}
L_{AV\phi} &=& g_{AV\phi}A_{\mu}\left[(p_{\phi}\cdot p_{V})g^{\mu\nu}
-p^{\nu}_{\phi}p^{\mu}_{V} \right]V_{\nu}
\end{eqnarray}
where $A, V$, and $\phi$ are respectively, axial-vector, vector and 
pseudoscalar fields.  A treatment of this interaction from a chiral lagrangian
approach has yielded results for photon production roughly consistent with 
those from this less complicated parametrization\cite{lxesgb,song}.
The coupling constant is adjusted to give a decay
$\Gamma_{a_{1}\to\pi\rho}$ = 400 MeV\cite{khphotons94} to match the
Particle Data Group\cite{pdg} value and the coupling of $a_{1}$ to 
$\pi$ and $\gamma$ is then taken to be the vector-dominance value of 
$g_{a_{1}\pi\rho}(e/f_{\rho})$, where $f_{\rho}$ is the coupling of rho to
pions.   Numerically these turn out to be $g_{a_{1}\pi\rho}$ = 16.1 
GeV$^{-1}$ and $g_{a_{1}\pi\gamma}$ = 0.81 GeV$^{-1}$ for $m_{a_{1}}$
= 1230 MeV.  The resulting prediction for the radiative decay width
$\Gamma_{a_{1}\to\pi\gamma}$ is 1.9 MeV, which is somewhat larger
than the 0.640$\pm$ 0.246 MeV obtained from a sole measurement\cite{mz84}.
Consequently, rates for photon or dilepton production from $\pi +\rho$ 
scattering through the $a_{1}$ might be corresponingly lower than
this approach would indicate.

The invariant rate for thermally producing a lepton pair of mass $M$ and
individual three momenta $\vec{\ell}_{+}$ and $\vec{\ell}_{-}$ via
the process $\rho(p_{a})+\pi(p_{b}) \to \pi(p_{1})+\ell_{+}\,\ell_{-}$
can be written generally as
\begin{eqnarray}
E_{+}E_{-} {dN\over d^{4}x\,dM^{2}\,d^{3}\ell_{+}\,d^{3}\ell_{-}} &=& 
{{\scriptstyle \cal N}\over 4(2\pi)^{2}} 
d\omega_{a}\,
d\omega_{b}\,d\omega_{1}
f(E_{a})f(E_{b})\tilde{f}(E_{1})
|\,\overline{\cal M}\,|^{2}\nonumber\\
&\ &\times\delta^{4}\left(p_{a}+p_{b}-p_{1}-\ell_{+}-\ell_{-}\right)
\delta\left[(\ell_{+}+\ell_{-})^{2}-M^{2}\right]
\label{eq:rate}
\end{eqnarray}
where $\scriptstyle \cal N$ is an overall degeneracy factor,
$d\omega_{a} = d^{3}p_{a}/[(2\pi)^{3}2E_{a}]$ and so on, $f$ is
the Bose-Einstein distribution, $\tilde{f}=1+f$ to account
for medium (Bose) enhancements and $|\,\overline{\cal M}\,|^{2}$ is
the initial spin averaged and final spin summed squared matrix
element. Integration over the full
individual lepton three momenta can be additionally performed to 
arrive at the total rate for given mass.  However, the differential
rate is more useful here since limited transverse momentum
and rapidity ranges can then be integrated to approximate experimental 
configurations.  

If we ignore the Bose enhancement then the rate can be simplified
as 
\begin{eqnarray}
E_{+}E_{-} {dN\over d^{4}x\,dM^{2}\,d^{3}\ell_{+}\,d^{3}\ell_{-}} &=& 
{{\scriptstyle \cal N}\over 16\pi^{4}}\,dz\,K_{1}(z)\,T^{2}
\lambda(s,m_{a}^{2},m_{b}^{2})\left[
E_{+}E_{-} {d\sigma\over dM^{2}\,d^{3}\ell_{+}\,d^{3}\ell_{-}}  \right]
\label{eq:rate2}
\end{eqnarray}
with $z=\sqrt{s}/T$ and $K_{1}$ is the first modified Bessel function.
For completeness, the cross section in brackets is 
\begin{eqnarray}
E_{+}E_{-}{d\sigma\over dM^{2}\,d^{3}\ell_{+}\,d^{3}\ell_{-}} &=&
{|\,\overline{\cal M}\,|^{2}\over 8(2\pi)^{5}
\lambda^{1/2}(s,m_{a}^{2},m_{b}^{2})}\delta\left((p_{a}+p_{b}-\ell_{+}-
\ell_{-})^{2}-m_{1}^{2}\right)\nonumber\\
&\ & \quad\times\delta\left[(\ell_{+}+\ell_{-})^{2}-M^{2}\right].
\end{eqnarray}
First, we integrate Eq.~(\ref{eq:rate2}) to get the 
total rate $dN/d^{4}xdM^{2}$.  To faciliate comparison with 
other rates from the literature we temporarily 
ignore collision broadening and set the pion chemical potential to zero.
In Fig.~\ref{fig:rate} the rates from $\pi\pi$ annihilation 
are compared with $\pi\rho\to\pi e^{+}e^{-}$.  
Formulas for annihilation are not shown because they are by now
textbook expressions\cite{wong} and have been recently used in
a similar study\cite{ssgale}.
For additional comparison we show also an estimate of pion scattering with 
bremsstrahlung\cite{kh92}.  All three contributions become roughly the
same around 350 MeV mass, which is the lower end of the window of the
observed excess over hadronic decays.  The mechanisms have different 
$p_{T}$ 
distributions for the leptons which will become important when kinematic
restrictions are imposed.  At the rho mass the two-pion annihilation 
is a factor of four above the $\pi\rho$ scattering.  We can 
expect this number since it is precisely the ratio of thermal production 
rates of neutral rho mesons via the two mechanisms
$\pi^{+}\pi^{-}\to \rho^{0}$ and $a_{1}\to \pi\rho^{0}$
\begin{eqnarray}
R &=& {dN^{\pi\pi\to\rho^{0}}\over d^{4}x}\left/
{dN^{a_{1}\to\pi\rho^{0}}\over d^{4}x}\right.
\end{eqnarray}
at $T$ = 160 MeV. 

In order to compare with experiment an integration of these rates over
a spacetime evolution must be preformed.  We assume for utmost simplicity
boost-invariant expansion\cite{bj} and neglect any transverse flow
effects.  This will provide a first estimate only while the need for
dynamic model calculations is duely noted.  Choice of
specific times or temperatures must be made.  Here we take 
$T_{i}=188$ MeV, $T_{c}=160$ MeV, $T_{f}=140$ MeV and $\mu_{\pi}=135$ MeV.
The rho is again assigned a collision broadened width. Total yield is the sum 
of contributions from the mixed plus cooling phases\cite{kk86}.  Written 
out in detail, it is
\begin{eqnarray}
{dN\over dy\,dM} &=&  {\pi R_{T}^{2}\over 2}\left(T_{i}\over T_{c}\right)^{6}
\tau_{i}^{2}r(r-1)
\int\,{d^{3}\ell_{+}\over E_{+}}{d^{3}\ell_{-}\over E_{-}}
\left[E_{+}E_{-}{dN\over d^{4}x\,dM\,
d^{3}\ell_{+}d^{3}\ell_{-}}(T=T_{c})\right] \nonumber\\
&\ & +3\pi R_{T}^{2}T_{i}^{6}\tau_{i}^{2}r^{2}
\int\limits_{T_{f}}^{T_{c}}
{dT\over T^{7}}\,{d^{3}\ell_{+}\over E_{+}}{d^{3}\ell_{-}\over E_{-}}\,
\left[E_{+}E_{-}{dN\over d^{4}x\,dM\,d^{3}\ell_{+}d^{3}\ell_{-}}
\right],
\label{eq:yield}
\end{eqnarray}
where $r$ is the ratio of degrees of freedom in QGP phase to that of 
hadron phase and is $\sim$ 12.  The pseudorapidity density will be assumed 
to be equal to the rapidity density $dN/dydM \approx dN/d\eta dM$.

Attempting to resemble the experiment as closely as possible lepton
transverse momentum integrations are limited to $p_{T}>$ 200 MeV/$c$ 
and pseuodrapidities by 2.1 $< \eta <$ 2.65.  There is also a cut in 
the experiment on angular separation for the pairs $\Theta>$ 35 mrad to 
avoid resolution difficulties.  We ignore this limitation here as it 
should only have small influence on the result for masses greater than
200 MeV.   Finally, we normalize just as in the experiment by
dividing the yield $d^{2}N/d\eta\,dM$ by the average charged particle 
pseudorapidity density $\langle dN_{\rm charged}/d\eta\rangle$ = 125. 

Having established the ranges kinematics ``visible'' by the experiment,
the rate from Eq.~(\ref{eq:rate}) can be obtained and then the convolution
of Eq.~(\ref{eq:yield}) performed.  The resulting normalized yield after
performing a mass-resolution smearing might
then be directly compared with the distribution observed by the
{\em CERES} collaboration.  Partial and total yield estimates are
shown in Fig.~\ref{fig:yield} as dashed ($\pi\pi$ annihilation), 
dot-dashed ($\pi\rho$ scattering), dotted (hadronic decays as presented
in Ref.\cite{ga95}) and solid (total of annihilation, scattering and decays).
Inclusion of the $a_{1}$ seems to account for part of the
excess but still leaves the door open for other possibilities.  
Additional pairs could be coming from $\pi\pi\to\rho\gamma^{*}$ and
$\pi\pi\to\pi\gamma^{*}$\cite{khseattle}.  They could also be coming 
from decays not already considered and possibly even a bremsstrahlung 
component\cite{dinesh}.

As a brief summary, we have pointed out that the recently observed dielectron
signal being in excess of hadronic decays in S+Au collisions at 200 GeV/u is
likely coming not only from $\pi\pi$ annihilation as was suggested, but
could also be partly due to pion plus resonance $\pi\rho\to\pi e^{+}e^{-}$ 
scattering.  The picture emerging for thermal production of low-mass 
dielectrons becomes richer and more challenging as it calls 
for next-to-leading order treatments of the photon self-energy, i.e.
$2\to 3$ body reactions (including the leptons).  This is not too 
surprising however, as there are many hadronic reactions of type
$a+b\to c+\ell^{+}\ell^{-}$ for the full set of light mesons.
The present result for $\pi\rho$ scattering suggests 
that moving towards quantitative interpretation of the {\em CERES} data 
will require consideration of such processes.

\section*{Acknowledgments}

The author would like to thank C. Gale and V. Koch for discussions.
This work was supported by the National Science Foundation under grant 
number PHY-9403666.

\begin{figure}
\caption{Dominant resonant contribution to 
$\pi\rho\to \pi\, e^{+}e^{-}$ scattering.}
\label{fig:feynman}
\end{figure}
\begin{figure}
\caption{Thermal rate for producing electron pairs at $T=160$ MeV for
the process $(\pi\rho\to\pi\,e^{+}e^{-})$ in the solid
curve, pion bremsstrahlung in dotted curve and $\pi\pi$ annihilation
in the dashed curve.}
\label{fig:rate}
\end{figure}
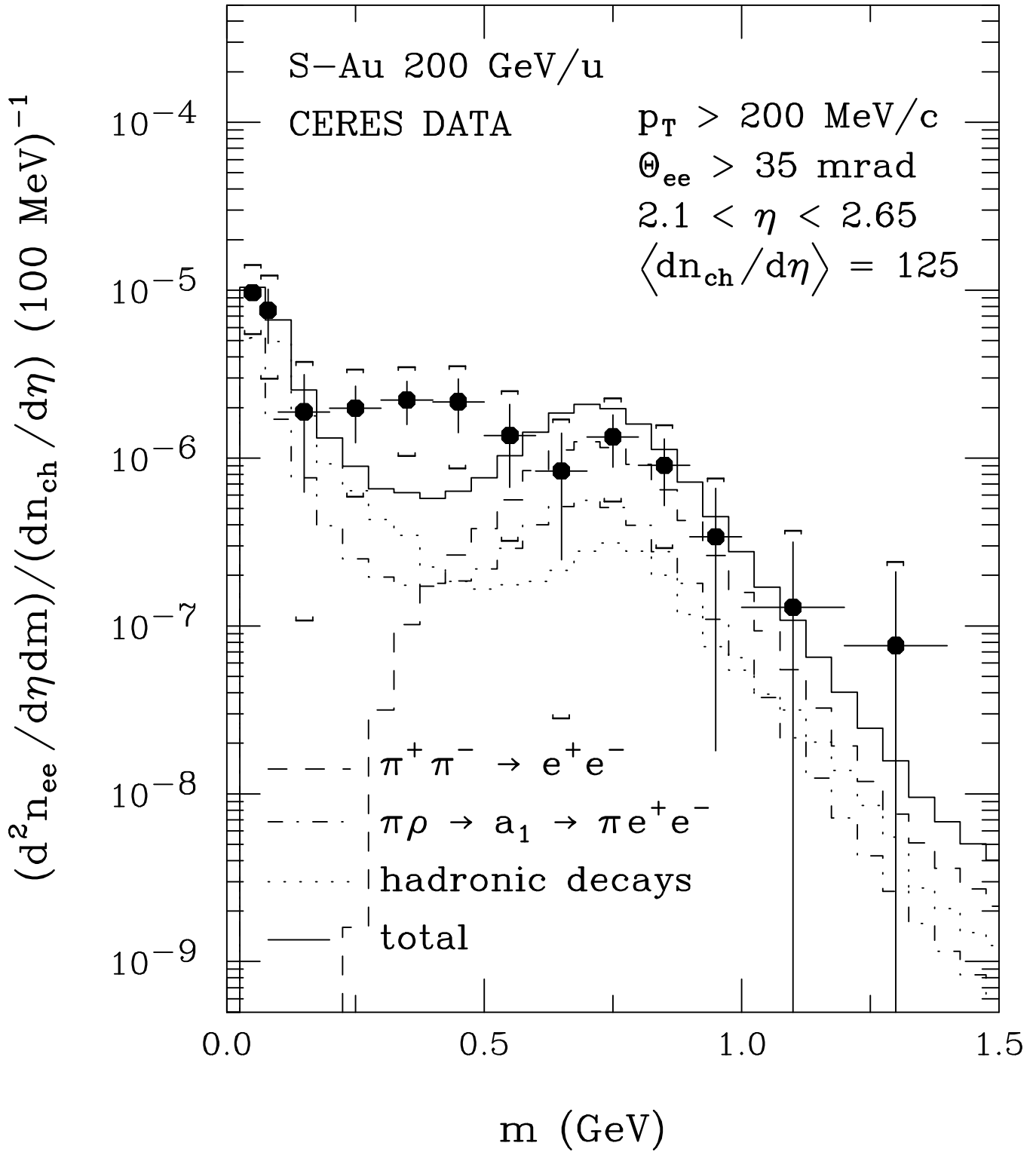
\begin{figure}
\caption{Normalized lepton pair yields observed in experiment by 
the {\em CERES} collaboration as compared with pion+resonance scattering
($\pi+\rho\to \pi\, e^{+}e^{-}$) in the dot-dashed distribution, 
estimated hadronic decays through event generation taken from Ref. [6]
(dotted), $\pi\pi$ annihilation in the (dashed)
and with the sum of annihilation, scattering and decays (solid).}
\label{fig:yield}
\end{figure}

\begin{references}
%
\bibitem[*]{myemail} electronic address: haglin@nscl.nscl.msu.edu
%
\bibitem{es80}E. Shuryak, Phys. Lett. {\bf 79B}, 135 (1978); Phys. Rep.
{\bf 67}, 71 (1980).
\bibitem{hwa85}R. Hwa and K. Kajantie, Phys. Rev. D {\bf 32}, 1109 (1985).
\bibitem{kk86}K. Kajantie {\em et al.}, Phys. Rev. D {\bf 34}, 2746 (1986).
\bibitem{jcl91}J. Cleymans, K. Redlich and H. Satz, Z. Phys. C {\bf 53}, 
517 (1991).
\bibitem{kh92}K. Haglin, C. Gale and V. Emel'yanov, Phys. Rev D {\bf 46},
4082 (1992); Phys. Rev. D {\bf 47}, 973 (1993).
\bibitem{ga95}G. Agakichiev {\em et al.}, Phys. Rev. Lett. {\bf 75}, 1272 
(1995).
\bibitem{li95}G. Q. Li, C. M. Ko and G. E. Brown, Phys. Rev. Lett. {\bf 75}
4007 (1995).
\bibitem{koch95}C. Song, V. Koch, S. H. Lee and C. M. Ko, 
Phys. Lett. B {\bf 366}, 379 (1996).
\bibitem{brown}G. E. Brown and M. Rho, Phys. Rev. Lett. {\bf 66}, 2720 (1991);
C. Adami and G. E. Brown, Phys. Rep. {\bf 224}, 1 (1993).
\bibitem{pisarski}R. Pisarski,  Phys.Rev. D {\bf 52}, 3773 (1995); 
hep-ph/9505257;
\bibitem{christ}N. H. Christ, Nucl. Phys. {\bf A544}, 81c (1992).
\bibitem{cgpl94}C. Gale and P. Lichard, Phys. Rev. D {\bf 49}, 3338 (1994).
\bibitem{kapusta}J. Kapusta, D. Kharzeev and L. McLerran, hep-ph/9507343.
\bibitem{lxesgb}L. Xiong, E. Shuryak and G. E. Brown, Phys. Rev. D {\bf 46},
3798 (1992).
\bibitem{song}C. Song, Phys. Rev. C {\bf 47}, 2861 (1993).
\bibitem{weldon93}H. A. Weldon, Ann. Phys. {\bf 228}, 43 (1993).
\bibitem{weldon}H. A. Weldon, Phys. Rev. D {\bf 28}, 2007 (1983).
\bibitem{jk}J. I. Kapusta, {\em Finite-Temperature Field
Theory}, Cambridge University Press, 1989.
\bibitem{khcoll95}K. Haglin, Nucl. Phys. A {\bf 584}, 719 (1995).
\bibitem{forthcoming}K. Haglin, J. Zhang and C. Gale, to be published.
\bibitem{khphotons94}K. Haglin, Phys. Rev. C {\bf 50}, 1688 (1994).
\bibitem{pdg}Review of Particle Properties, Phys. Rev. D {\bf 45-II} 
(1992).
\bibitem{mz84}M. Zielinski {\em et al.}, Phys. Rev. Lett. {\bf 52}, 1195
(1984).
\bibitem{wong}C.-Y. Wong, {\em Introduction to High-Energy Heavy-Ion 
Collisions}, World Scientific, pg. 332 (1994).
\bibitem{ssgale}D. K. Srivastava, B. Sinha and C. Gale, Phys. Rev. C
{\bf 53} R567 (1996).
\bibitem{bj}J. D. Bjorken, Phys. Rev. D {\bf 27}, 140 (1983).
\bibitem{khseattle}K. Haglin, Talk given at the INT/RHIC Workshop
``Electromagnetic Probes of Quark Gluon Plasma'', INT, Seattle
24--27 January 1996.
\bibitem{dinesh}D. Pal, K. Haglin and D. K. Srivastava, submitted to 
Phys. Rev. D.
\end{references}
\end{document}